# Analyzing the Influence of Processor Speed and Clock Speed on Remaining Useful Life Estimation of Software Systems

Mohammad Rubyet Islam[1] and Peter Sanborn[1]

[1] University of Maryland, College Park, MD, 20742, USA
`rubyet@terpconnect.umd.edu`

**Abstract.** Prognostics and Health Management (PHM) is a discipline focused on predicting the point at which systems or components will cease to perform as intended, typically measured as Remaining Useful Life (RUL). RUL serves as a vital decision-making tool for contingency planning, guiding the timing and nature of system maintenance. Historically, PHM has primarily been applied to hardware systems, with its application to software only recently explored. In a recent study we introduced a methodology and demonstrated how changes in software can impact the RUL of software. However, in practical software development, real-time performance is also influenced by various environmental attributes, including operating systems, clock speed, processor performance, RAM, machine core count and others. This research extends the analysis to assess how changes in environmental attributes, such as operating system and clock speed, affect RUL estimation in software. Findings are rigorously validated using real performance data from controlled test beds and compared with predictive model-generated data. Statistical validation, including regression analysis, supports the credibility of the results. The controlled test bed environment replicates and validates faults from real applications, ensuring a standardized assessment platform. This exploration yields actionable knowledge for software maintenance and optimization strategies, addressing a significant gap in the field of software health management.

## 1 Introduction

This research paper pioneers in bridging a critical gap in the realm of software system management by applying Prognostic and Health Management (PHM) and Remaining Useful Life (RUL) concepts to software systems. Despite extensive research on software defect prediction, reliability, and predictive maintenance, the application of PHM in software has been notably underexplored. Our study introduces a novel methodology, integrating dynamic PHM principles with software components, especially focusing on version upgrades. We innovatively utilized embedded sensors, health monitoring tools, and user feedback to calculate the RUL of software systems. In this venture, we merged methodological advancement with data science techniques, aiming for an automated classification of software faults and enhancement requests. The approach evolves over time, employing data mining and machine learning techniques, including Natural Language Processing (NLP) for classifying user stories and sizing them. We employed k-means clustering for grouping similar releases, which aids in selecting the most fitting prognostic model, considering various factors like data type and volume, variable relationships, and environmental influences. Significantly, this research addresses the considerable financial impact of poor software quality, evidenced by the $2.08 trillion cost in the USA in 2020 (Krasner,2021), underscoring the urgent need for enhanced software failure prediction and management. Unlike existing diagnostic models that predominantly focus on identifying current faults based on historical data, our approach innovatively predicts the future states and RUL of software. This predictive capability is essential for effective maintenance and resource planning. By overcoming the limitations of traditional diagnostic models, our research proposes a more efficient prognostic methodology. This advancement is poised to revolutionize software health management, resulting in more reliable software, reduced maintenance costs, and improved decision-making in software lifecycle management.

## 1.1 Prognostic and health management

Prognostics and health management (PHM) methods reduce time and costs for the maintenance of products or processes through efficient and cost-effective diagnostic and prognostic activities. PHM systems use real-time and historical state information of subsystems and components to provide actionable information, enabling intelligent decision-making for improved performance, safety, reliability, and maintainability. The goal of PHM is to provide decision support; that is, actionable information to support system support decision making. While diagnostics is the process of detection and isolation of faults or failures, prognostics is the process of prediction of the future state or remaining useful life (RUL) based on current and/or historic conditions [1]. Prognostics is based upon the understanding that systems/products fail after a period of degradation, which if measured (via an RUL), can be used to prevent system breakdown, avoid collateral damage, and minimize operation costs [2]. The inability of the system to perform its intended function is most often a failure beyond which the system can no longer be used. The predicted time-to-failure is referred to as the remaining useful life (RUL) [3]. Remaining useful life (RUL) represents the useful life left in an asset at a particular time of operation. Its estimation is central to condition based maintenance and prognostics and health management. The uncertainties of the estimated system condition affect any future prediction [4,5]. Some failures are intermittent and difficult to predict [6]. Hence, there is no universally accepted methodology for prognostics [7]. Current prognostics technology is immature due to the lack of uncertainty calculations, validation and verification methods, and risk assessment for PHM system development [8]. However, the most important fact for this paper is that PHM has only been applied to hardware, not to software.

## 1.2 Software PHM and RUL

Software Prognostics and Health Management (SW PHM) can assist in decision-making regarding software version updates/upgrades, predicting Remaining Useful Life (RUL), software component updates/upgrades, and as part of the decision-making process for reengineering or abandoning legacy code. This is a relatively new research area, with no known research on the application of PHM to software systems. However, there is relevant research on diagnostic health management of software systems.

Technically, software code does not decay, nor do its individual components. However, software systems can degrade over release cycles. Software degradation, also known as software decay, aging, rotting, smell, degradation, code rot, bit rot, software entropy, and software erosion, represents a slow deterioration of software performance over time or its diminishing responsiveness, eventually leading to software becoming faulty, unusable, and in need of upgrades. SW degradation is not a physical phenomenon. Software failure corresponds to unexpected runtime behavior observed by a user of the software, whereas a fault is a static software characteristic that causes a failure to occur. Faults result from incorrect design choices, simple human blunders, or the forces of nature [9]. The term "bug" or "defect," sometimes used to describe a fault, error, or failure, is not precise enough [10]. For this paper, we will use "faults" and "failures" instead of "bugs/defects" to avoid confusion.

In the real world, response time (RT) is not affected only by software releases. Although the impacts of environmental variations in the test beds that could affect RT have been controlled, RTs can still be affected by other environmental attributes, such as clock speed, processor speed, RAM, and others, which influence the performance of software. Table 1 shows the estimated cost of software failures in 2018.

**Table 1.** Cost of software failures in 2018 [9].

| Number of Failures | 606 from 314 companies |
|---|---|
| Financial Loss | US $1.7 trillion |
| People Affected | 3.6 billion |
| Lost to Downtime | 268 years |

This conference paper, organized into six sections, delves into the innovative application of Prognostic and Health Management (PHM) models for estimating the Remaining Useful Life (RUL) of software systems. Section 2 lays the groundwork by reviewing existing research and highlighting the unique aspects of PHM in software compared to hardware. Section 3 introduces a novel, fusion-based approach for prognostic estimation of software RUL. This approach is practically demonstrated in Section 4 using an open-source software system. The results and insights gained from this application are thoroughly discussed in Section 5. Finally, Section 6 wraps up the paper, summarizing key findings and pointing out potential avenues for future research in this emerging field.

## 2    Literature Review

This section reviews relevant literature associated with software reliability prediction, software predictive maintenance, software decay, software performance prediction, and software defect prediction. Although all these topics are diagnostic in nature (as opposed to prognostic) and do not generate Remaining Useful Life (RUL) estimates, they are nonetheless relevant to the application of PHM to software systems. According to ISO 9126, reliability is the capability of the software product to maintain a specified level of performance when used under specified conditions [10-14]. However, reliability does not predict remaining useful life (RUL). PHM and RUL can be used to identify reliability over release cycles, and accurate predictions of reliability may avoid failures and significant disruptions [7]. Thus, increasing a system's reliability (time or cycles between failures). PHM can assist in making accurate predictions in a dynamic manner. PHM is a methodology for evaluating the reliability of a system to predict and mitigate failures [6]. Software maintenance is the process of modifying a software system or component after delivery to correct faults, improve performance, or adapt to a changed environment [15]. PHM can be used as a tool to plan predictive maintenance activities. Software Defect Prediction (SDP) predicts defects in modules that are defect-prone and require extensive testing [16-17]. However, defect prediction is a diagnostic process based on historical data. Defect prediction does not account for enhancement requests and does not predict remaining useful life (RUL). Defect predictions are mainly conducted during the software development process and are less common in the production environment. Defect prediction methods can be used as elements of RUL estimation. Integrated Software Health Management (ISWHM) [18] is a health monitoring system that monitors onboard software and sensors to detect failures in both hardware and software components using probabilistic modeling. ISWHM is a diagnostic approach that does not address future enhancement requests. However, methods used in ISWHM to collect fault data can be utilized in data collection for PHM. Models to evaluate software decay [19,20] are used to understand software degradation mechanisms. Such knowledge can assist in calculating PHM-based RUL.Albu and Popentiu-Vladicescu attempted to predict response time for web services and web applications [21]. They tried to measure response time/execution time at a specified point in time based on the existing system; however, they did not consider the influences of faults and enhancement requests on performance parameters. PHM potentially addresses this gap.

The existing body of literature on software management reveals significant gaps, particularly in the realm of addressing the Remaining Useful Life (RUL) of software systems. Current research in areas such as software reliability, predictive maintenance, decay, performance prediction, and defect prediction predominantly adopt a diagnostic approach, focusing on present conditions rather than prognostic aspects like predicting future states or RUL. For example, although ISO 9126 and related works (10-14) discuss software reliability, they do not extend their scope to RUL prediction. Similarly, methods in Software Defect Prediction (16-17) and Integrated Software Health Management (18) provide insights into immediate software issues but fall short in forecasting future enhancements or ensuring long-term viability. Even attempts to measure performance parameters, such as response times in web services (21), overlook the impacts of faults and enhancement requests. This paper aims to address these shortcomings by proposing a PHM-based methodology that surpasses diagnostics and offers a predictive view of software health. It effectively fills the gap in current literature by providing a comprehensive approach to estimating RUL, considering various factors that affect software longevity and performance.

## 3    Proposed Solution

This section describes the methodology developed for predicting the Remaining Useful Life (RUL) of a software application. It also explains how two selected environmental parameters, such as clock speed and processor speed, impact the estimation of the remaining useful life (RUL) of a software system. In this section, we briefly discuss the method for prognostically and continuously predicting the RUL of a software (SW) system based on usage parameters (e.g., the number and categories of releases) and multiple performance parameters (e.g., response time) to provide background understanding. In this paper, we measured correlations between the predictors and the predictive variables to understand their relationship. A higher correlation among the variables (discussed in the following section) confirmed the existence of a linear relationship between the predictors and the predictive variables. Hence, a linear regression model was selected to predict the performance parameters. In the following subsection, the adjusted RUL is measured based on test beds with varying operating systems (32-bit versus 64-bit) and by changing clock speeds. The model was validated in two different ways. The first one is data-driven validation, where actual data (on performance parameters) generated by the test beds were compared with predicted data. The second one involves measuring the fitness of fit using the adjusted R2. A low p-value (<0.05) indicated that we can reject the null hypothesis, i.e., the changes in the predictor (input) variable are associated with the changes in the predictive (output) variable. The test beds produced response time (RT) over multiple releases in a controlled and standard test (staging) environment, based on user-reported and system-generated (through sensor/health monitoring tools) faults and enhancement requests.

We continuously collected data on SW enhancement requests and SW faults from the open-source Bugzilla repository, caused by software component updates, upgrades, and integrations, as well as changes in environmental variables (OS and clock speed) [22]. Users collect such data based on their experiences, enhancement requirements, and data generated by SW sensors and health monitoring tools (e.g., Nagios Core, Splunk). Since faults and enhancement requests are mostly written in descriptive format, most of the raw data are text data (unstructured). However, other data types also exist, including continuous data (structured) and categorical data. For this exercise, we selected data that have an impact on the target performance parameter, which is response time in this case. We also collected readings on response time (RT) for individual versions of Bugzilla from a standard test bed. Our test bed specifications were as follows:

Microsoft VMWare (Windows 10)

OS: 64-bit and 32-bit

Clock Speed: 1.8 GHz and 2.4 GHz

RAM: 4 GB

HDD: 50 GB

The impacts of the faults/enhancements were calculated by estimating story points and factoring them with impact factors (also known as severity). Estimation of story points was carried out based on Fibonacci sequences [23], where the sequence used is 1, 2, 3, 5, 8. Here, 1 represents the smallest story point, and 8 represents the largest. Story points include attributes like the volume of work, inherent uncertainty, complexity, and efforts required to complete each fault/enhancement request. However, higher story points do not necessarily mean higher impacts on the target performance parameter. Hence, we introduced impact factors, interpreted on a scale. Table 2 is the custom-built lookup table for conversions between impact scales and impact factors.

**Table 2.** Impact factor table – conversion between impact scales and numeric factors [24]

| Impact Scale | Impact factor |
|--------------|---------------|
| Critical | 1 |
| Major | 0.75 |
| Medium | 0.5 |
| Minor | 0.25 |

The higher the impact, the greater the impact factors. The product of the impact factors and the story points results in a weight factor (WF). This WF serves as the predictive (input) variable used to estimate the predictor, response time (RT).

Initially, we classified the faults and enhancements into custom categories, which is the first step in analyzing such data. We employed NLP (Naïve Bayes) for this classification [25]. Classification also aids in estimating story points and impact factors. We then formed clusters of analogous releases based on fault data and enhancement requests using k-means clustering, an unsupervised clustering algorithm that self-learns, evolves, and can identify new clusters for new releases. Based on the estimated WF, we calculated the Remaining Useful Life (RUL). RUL changes over time as the number of faults/requirements changes. Therefore, we refer to our model(s) as dynamic. The calculated dynamic RUL can be valuable for decision-making in SW upgrade/update/integration, addition of new features, improving existing features, rejuvenation, system re-engineering, predictive maintenance, budgeting, and other areas.

There are no industry standards for response time (RT) thresholds. Thresholds are determined based on de facto standards and vary depending on the type of application, user requirements, and other factors. However, there are three main time limits to keep in mind when optimizing web performance [26]:

0.1 second: The time for user to perceive that system is reacting instantaneously.
1.0 second: The limit for the user's flow of thought.
10 seconds: The limit for maintaining the user's attention.

In this paper, we consider 10 seconds as the threshold. We also focus on only one performance parameter to demonstrate this methodology with respect to time and resources. However, a similar method could be applicable when multiple performance metrics are considered. To do so, we need to identify the input variables (enhancement requests, faults, and other relevant parameters) that affect the other performance parameters to be considered. Then, we will identify the RUL in terms of the new performance parameters. Producing such RUL values for individual performance parameters is beneficial for understanding to what extent reported faults or change requests affect different performance parameters. This way, users can make tradeoffs regarding which changes to implement based on their priorities for performance parameters.

To analyze text data, we utilized free versions of Google Colab Notebook with Python as the scripting language for clustering and classifications. Google Colab provides faster processing capabilities with higher computing power, which is essential when processing text data. For predicting response time (RT) based on regression analysis, we used R-Studio.

The equations to calculate PV and CPV for each software releases are:

$$PV_1 = (\pm SP_1)(IF_1) + (\pm SP_2)(IF_2) + \ldots + (\pm SP_n)(IF_n) \qquad (1)$$

$$CPV_m = PV_1 + PV_2 + \ldots + PV_m \qquad (2)$$

Where, CPV is the cumulated predictive variable, *PV* is the total story point for individual releases (that can either be positive or negative, see Section 3.5), *SP* is the estimated story for individual fault/enhancement, *IF* is the impact factor, *n* is the number of faults and enhancements within a release, and *m* is the number of releases after cumulation. Equation (1) estimates the total story points, *PV*, of faults and enhancements by taking account of the impacts of individual faults/enhancements on performance parameters by factoring the story points with impact factors.

### 3.1     Performance and usage parameters

The following is a list of parameters that have been used in the industry to measure software performance [15-16,23,25,27]:

- **Execution time:** Time spent by the system executing the task [27].
- **Timeliness:** Timeliness is measured in terms of response time and throughput [15].
    - **Throughput:** The rate at which a system can process inputs. [23,25,27,28].
    - **Response time:** The time between the end user requesting a response from the application and a complete reply [27].
- **Availability:** The fraction of time an application is available to the end user [27].
- **Utilization:** The percentage of the theoretical capacity of a resource that is being used [27].

Unlike HW, SW generally does not decay over time. Hence, time is not an appropriate usage parameter. software rots/degrades/decays as we make changes and changes happen over releases. Hence, release cycle is a better fit when it comes to usage parameters.

### 3.2. Environmental Variables on RT

The environmental attributes that influence the performance of software. And they include:

- **Operating system (OS):** Operating system (OS) manages memory and processes, as well as software and hardware of a machine(computer). There are attributes of the OS that influence software performance. One of them is *bit*. Bit is an architecture type that represents the capacity of data transfer and is closely related to the processor type (32-bit Versus 64-bit).
- **Clock speed/processor speed:** The clock speed, also known as clock rate, or processor speed, indicates how fast the CPU can run. Clock speed measures the number of cycles the CPU can execute per second. Clock speed is measured in Hz (GHz, MHz, KHz).
- **RAM (Random Access Memory):** RAM is a computer's (also referred to as, machine) short-term memory, where data is stored temporarily so that it can access memory much faster than data on a hard disk, SSD, or other long-term storage devices.
- **Number of machine cores:** The core is a small central processing unit (CPU) or processor inside a big CPU or CPU socket. More cores mean a faster running machine.
- **Storage or hard drive space:** A hard drive is the hardware component that stores all the digital contents. Larger storage results in faster performance of the machine (Arpaci-Dusseau & Arpaci-Dusseau, 2014)

# 4 RUL estimation by accounting impacts of environmental variables

We have used different versions of Bugzilla [29] as test beds that were built upon a standardized and scaled staging environment on Microsoft Virtual Machines.

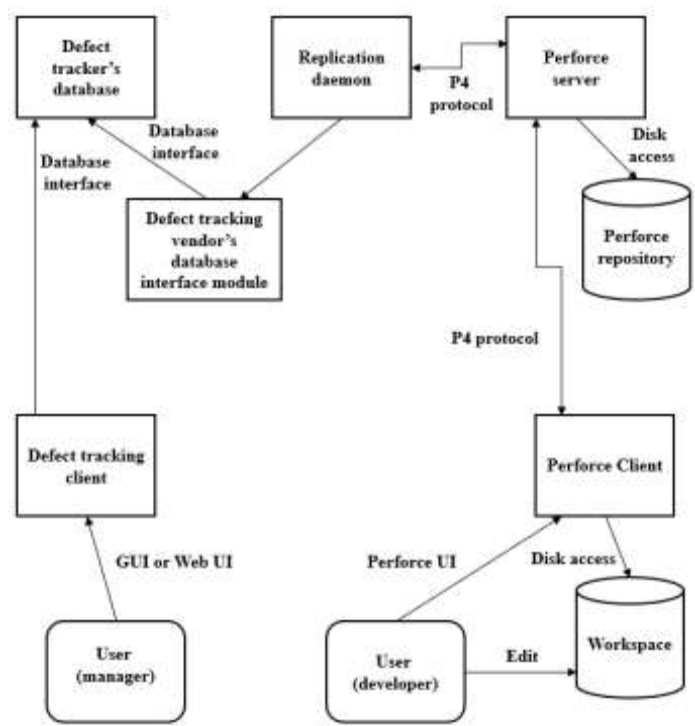

**Fig. 1.** Bugzilla architecture [26]

Bugzilla is an open-source tracking tool, used to track defects. It has multiple integrated modules including but not limited to Perl, Apache, SQL, PostgreSQL, Oracle, and SQLite (see Fig. 1.). The application and functionality of Bugzilla is irrelevant for this demonstration; Bugzilla's important attribute is that it has been in existence for 22 years and has an open-source defect tracking system. For demonstration purposes we will only focus on *response time* as a performance parameter and faults and enhancement requirements, and environmental variables that are related to response time only.

For this demonstration, we have collected Bugzilla data on SW faults and Enhancement Requests (Updates/Upgrades). The data types are [Update data types to address the 2 env param]:

- Unstructured data (texts): Faults & enhancement requests

- Categorical variables: Fault/enhancement classes
- Continuous variables: Story points, impact factors, weight factors (WF).

We conducted a prognosis of target variables based on predictive variables from analogous releases. Given the linear correlation between target variables and independent variables, we employed regression analysis for the prognosis of response time based on fault/enhancement weight factors (WF).

Initially, we utilized historical health monitoring and fault data collected by sensors, users, and software testers, as well as hardware and software specifications, to predict WF (Fig 2) and response time (Fig 3) and calculate RUL. We continued to collect fault data and new enhancement requests dynamically and fed them into the proposed hybrid model to continuously calculate RUL. Thus, we formed a dynamic fault and RUL prediction model.

We collected historical release information on ten different Bugzilla releases. In a standardized staging environment, we took multiple readings of average response time and computed an average of the averages. JMeter [30] was used to collect performance readings on response time.

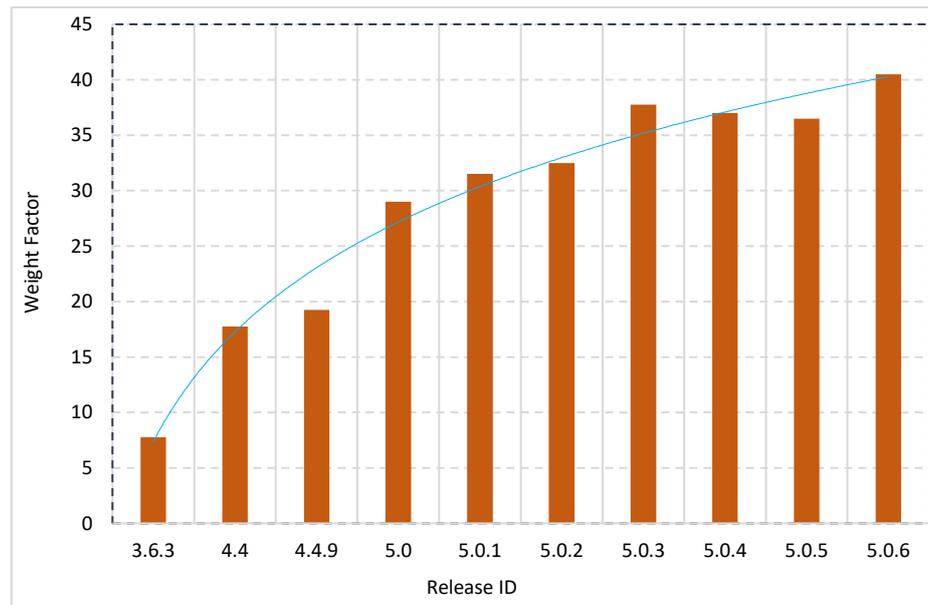

**Fig 2:** Weight factor (WF) over multiple bugzilla releases

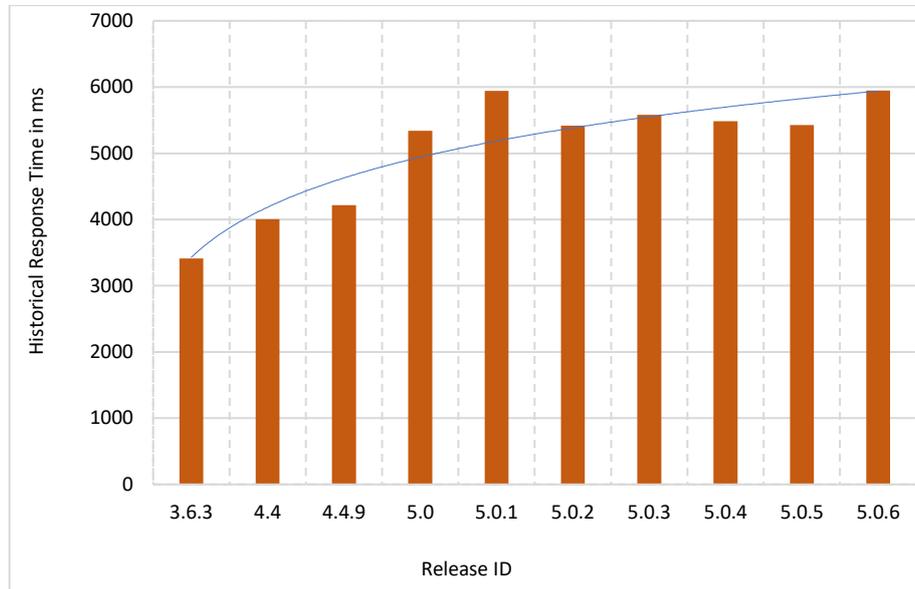

**Fig. 3.** Response time (RT) over multiple bugzilla releases

We also collected existing fault data and enhancement requests, reported in Bugzilla open-source repository, and reported by the Bugzilla user community. We used this data to plan for future releases. We also collected data on response time by varying environmental variables such as OS specifications and clock speed and keeping the other environmental variables the same.

## 5     RESULTS

After estimating story points for individual faults/ enhancements, we multiplied them with the impact factors to identify the final weight factors. To understand the relationship between the target variable (RT) and the independent variable (WF) we performed a correlation analysis (Fig. 4.). Quadrant I show a higher degree of association (0.95) between RT (Quadrant IV) and WF (Quadrant II). The direction in Quadrant III confirms the positive nature of this relationship, i.e., as weight factor increases, response time increases.

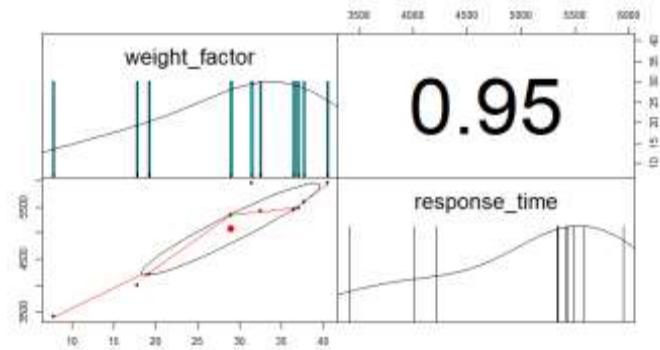

**Fig. 4.** Correlation analysis between WF and RT using R Studio

Given the high correlation factor we applied Linear Regression (LR) to estimate RT based on WF. Since all the fixes and enhancements get added to the next release, our weight factors are cumulative as well. We divided the historical data into tests (80%) and training (20%) data sets. After training the univariate linear regression with the training data set, we validated the model using test datasets and ran cross validation. The model is also validated statistically, with R-squared value and p-value. Following is the model performance result:

> Multiple R-squared value:0.90
> Adjusted R-squared: 0.88.
> p-value: 0.0004

Here, the R-squared value is acceptable with a p-value of 0.0004, which is significantly smaller than 0.05.

The next step was to identify the WFs of the currently reported faults and enhancement requests using the method stated earlier. Various combinations (4 in this case) of the same release contents were planned and spread over multiple releases. Using the trained model, we *predicted* response time (RT) and calculated the remaining useful life (RUL) after each release. As we can see in Fig. 5., multiple combinations of releases result in exceeding the RT threshold (the threshold is user defined) after various numbers of release cycles. Based on which we can determine which combinations of releases give us the best RT sustainability. In this case, Combo-1 gives the maximum number of releases before the SW system reaches the threshold.

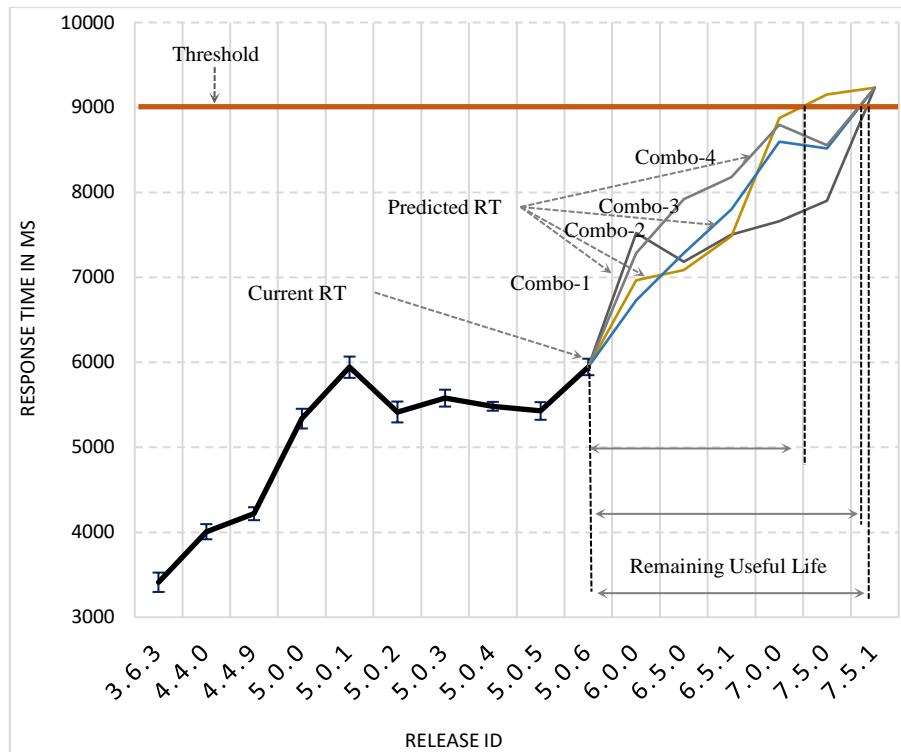

**Fig. 5.** Estimation of RUL of Bugzilla on 64-bit OS and 1.8 GHz clock speed

Following that, we estimated the impact of changes in the operating system (32-bit vs 64-bit) and clock speed.

**Operating System (64-bit versus 32-bit).** Experiements are carried out by varying the operating system but keeping all the other test bed attributes the same. Individual virtual environments are created for multiple versions of Bugzilla and the corresponding RTs are captured in Fig. 6. The host machine and virtual machine specifications are the same, as described earlier, shows the impacts of both 32-bit and 64-bit OS on RT. In both the cases the machine clock speed is 1.80 GHz. Experiemental results confirm the assumptions that RT is expected to be higher 32-bit OS comparing to to 64-bit OS for all releases, as the 64-bit OS has a higher processing speed. RT is also predicted for all future releases based on 32-bit OS for all the combos. Predicted RTs are plotted against release cycles to identify RUL both on 32-bit and 64-bit OS, as shown in Fig 7. This figure clearly shows that when Bugzilla is configured on a 32-bit machine, it reaches thresholds for all the combos even before Release 6.0.0. Whereas 64-bit OS extends the remaining useful life of Bugzilla for a few more releases.

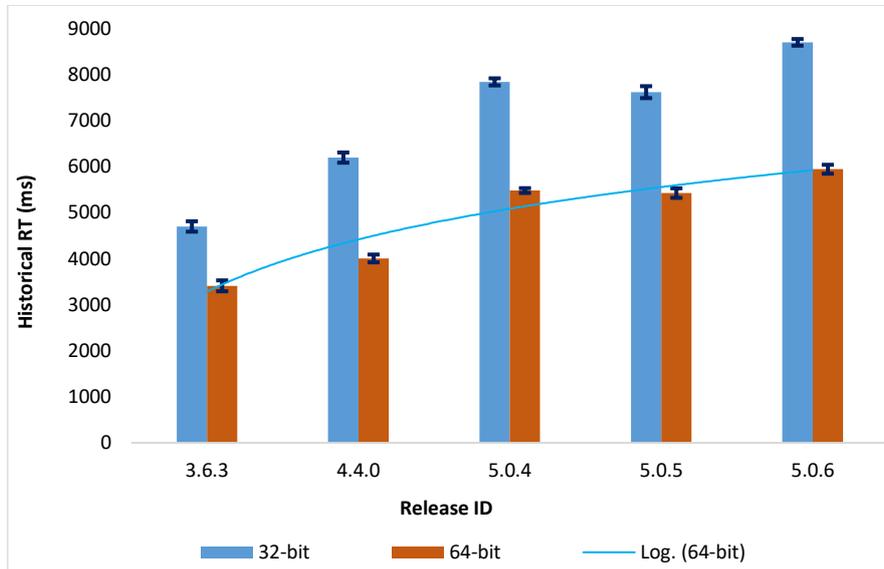

**Fig. 6.** RT on 1.80 GHz machine with 32-bit versus 64-bit OS

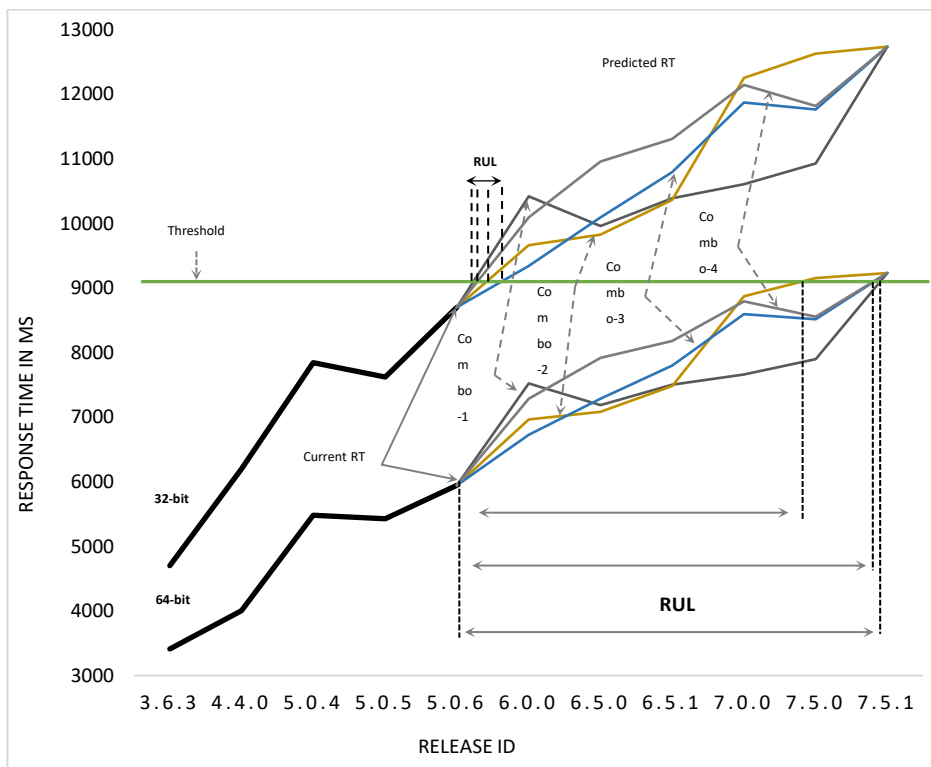

**Fig. 7**. Estimated RUL on 1.80 GHz machine with 32-bit versus 64-bit OS

**Clock Speed (Processor Speed).** As stated earlier, clock speed measures the number of cycles the CPU executes per second. During each cycle, billions of transistors within the processor open and close. Fig 8 shows a graphical representation of cock speed. And Fig 9 represents historical upgrades in clock speed over time (Microprocessor Chronology, 2022). According to this diagram, the highest speed of 5.5 GHz existed in the market by 2012. Processors that came out in the market beyond 2012 are either the same or lower than 5.5 GHz.

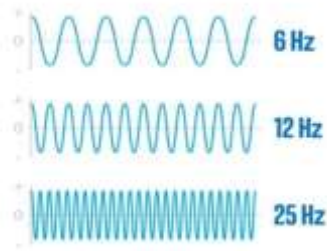

**Fig. 8**. Clock speed cycle (Intel, 2022)

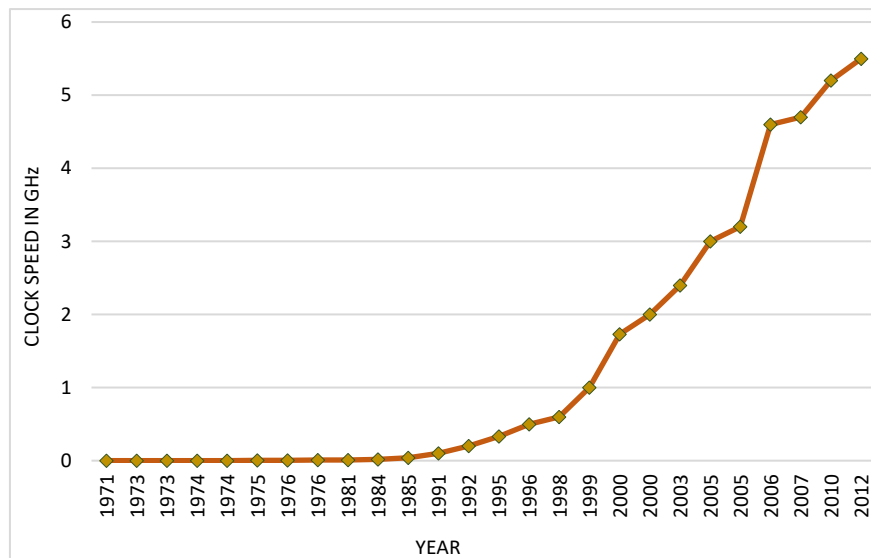

**Fig. 9.** Historical clock speed (Microprocessor Chronology, 2022)

The clock speed of the host machine on which the VMs were running, and the test beds were built is 1.80 GHz. This section estimates adjusted RULs based on varying clock speeds (2 GHz and 2.4 GHz, respectively). Experiments are carried out by varying clock speed of the test beds for Bugzilla 5.0.4 and RT is measured (Fig 10) for all the different clock speeds.

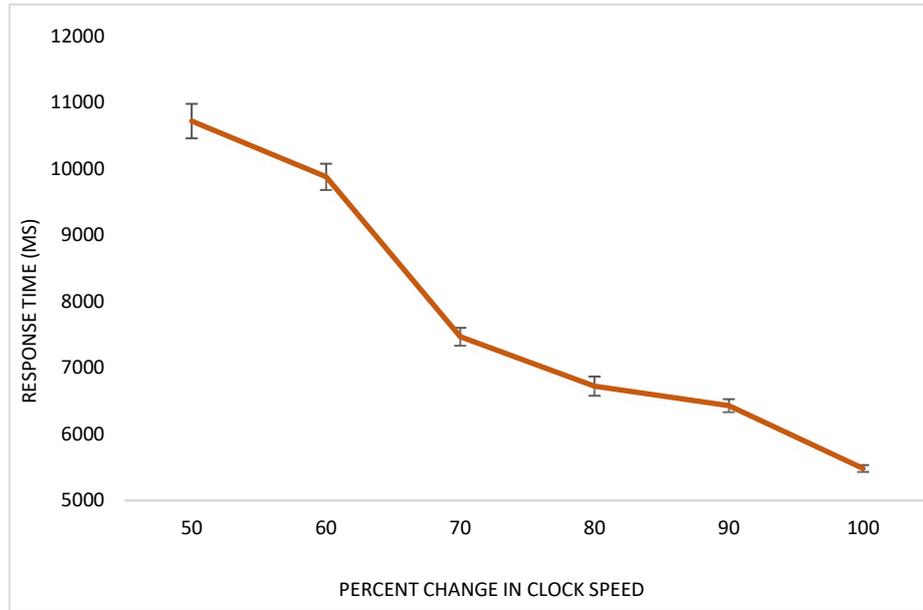

**Fig. 10.** Impacts of percent change in clock speed on RT

After aggregation, it is identified that for each 10% increase in clock speed a mean 12.27% decrease (an improvement) in RT is observed. This aggregated decrease is used as a factor to modify the predicted RTs for all future releases to adjust RTs. Equation (3) is used for this estimation. The adjusted RTs are then used to predict modified RULs.

$$RT_n = RT_o \left[ 1 - \left( \frac{0.1227}{0.1} \right) \left( \frac{Hz_n - Hz_o}{Hz_o} \right) \right] \qquad (3)$$

where, $RT_n$ is the adjusted RT, $RT_o$ is the previously predicted RT for future releases (Release 6.0.0 to Release 7.5.1), $Hz_n$ is the new clock speed and $Hz_o$ is the existing clock speed.

The first improvement in clock speed is made at Release 6.0.0 (2.0 GHz clock speed) and the second improvement is assumed to be made at Release 7.0.0 (2.4 GHz clock speed). As shown in Fig 11, the clock speed is upgraded from 1.80 GHz to 2.0 GHz at Release 6.0 and continued until Release 7.0.0. At Release 7.0.0, the processor is upgraded once more from 2.0 GHz to 2.4 GHz. Both the upgrades were made on 64-bit machine and resulted in a significant reduction in RTs that led to longer RUL.

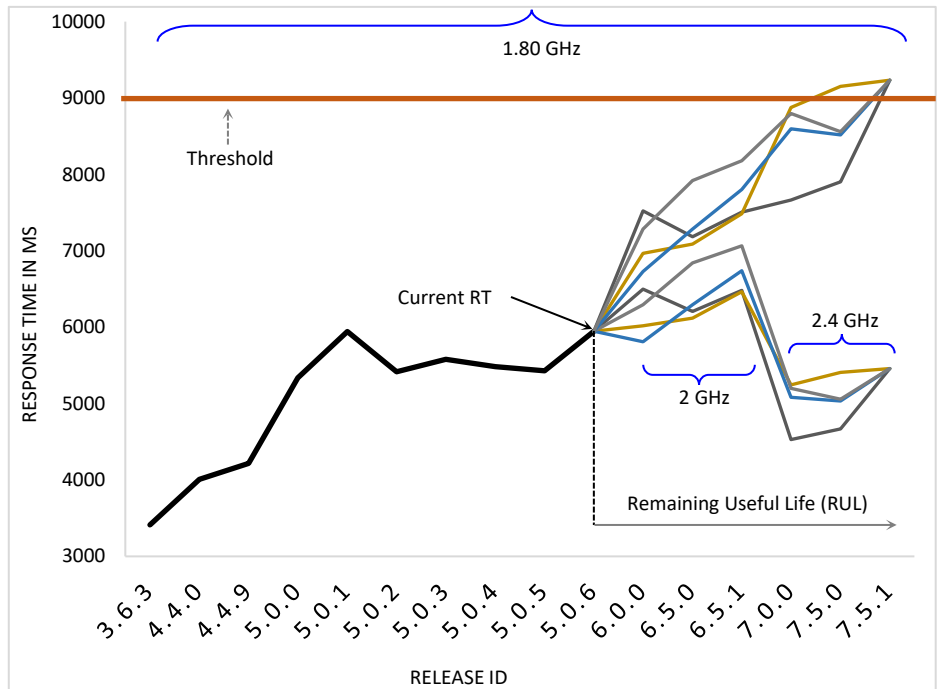

**Fig. 11.** Adjusted RUL after upgrades in clock speed on a 64-bit machine

Again, in Fig 12, RUL is estimated based on modified clock speed but on a 32-bit machine. Both 2.0 GHz and 2.4 GHz processors show better promises than RTs on 1.80 GHz processor on 32-bit machine. Although, in few combos, RTs still cross the threshold of 9 sec in Release 6.0.0 to Release 6.5.1 on 2.0 GHz processor, all the combinations of RTs are <9 sec once processor speed is increased to 2.6 GHz.

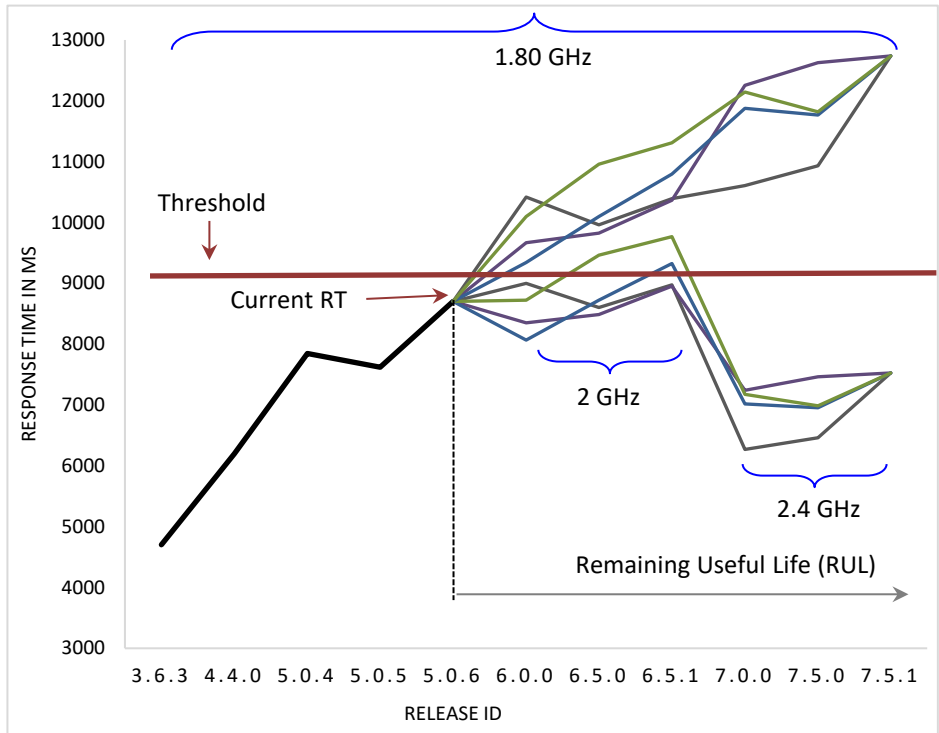

**Fig. 12.** Adjusted RUL after upgrades in clock speed on 32-bit machine

Furthermore, as shown in Fig 13, when the upgrade is made both in OS (from 32-bit to 64-bit) and clock speed (from 1.8 GHz to 2.0 GHz and 2.4 GHz, respectively) a significant improvement in RUL is observed across all the combinations of releases.

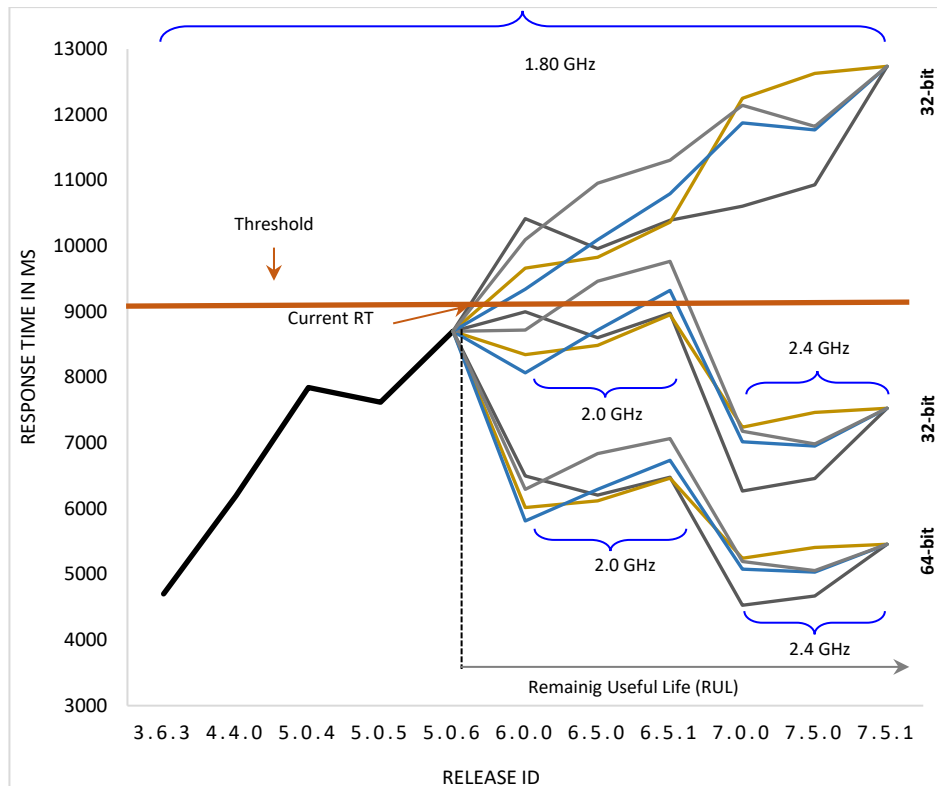

**Fig. 13.** Adjusted RUL after upgrade in GHz and bit (32 versus 64)

RUL estimations, such as those shown in above Figures give us control over release planning for software updates/upgrades that enables users to get the longest RUL for the corresponding software systems. We, also, can control and identify numbers of releases and the points at which we may need to do update, upgrade, rejuvenation, re-engineering, predictive maintenances, and others.

## 6 Conclusion

This paper demonstrated that Prognostic Health Management (PHM) technique can be utilized on SW system taking accounts of impacts such as version upgrades/updates in terms of changes in code and environmental attributes (e.g., 32-bit vs 64-bit CPU architecture, clock speed). This demonstration is based on one performance parameter, the response time (RT). The methodology described in this dissertation potentially applies to all software types that have performance requirements, especially RT. More research can be carried out to identify individual and collective impacts of other environmental paraments on RT and other performance parameters.

Future research in the field of Prognostics and Health Management (PHM)-based Remaining Useful Life (RUL) estimation for software systems is crucial, particularly in expanding beyond response time to include parameters such as execution time and availability. This requires the utilization of diverse health monitoring tools and data assessment methods. An area of utmost importance is the evaluation of software performance across different operating systems, considering factors like OS architecture and processor compatibility. Moreover, there is a clear need for more robust and data-driven validation methods for future software releases. To enhance the accuracy of prognostic models, it is essential to explore a wider range of predictive algorithms, including machine learning and ensemble methods. As the volume of data increases, it becomes increasingly important to refine the data quality assessment matrix. A deeper understanding of software performance can be gained by examining software release cycles and other usage parameters. To add sophistication to the process, automating the categorization of faults and enhancements, possibly through Natural Language Processing (NLP), and exploring diverse clustering algorithms would be beneficial. Additionally, establishing industry standards for software performance thresholds, adapting to user requirements, and considering user perspectives on software durability are key directions for future research. The validation scope can be broadened by applying these methods to various software systems like Bugzilla, particularly those with publicly available data. Lastly, incorporating additional hardware and software usage parameters, such as degradation and maintenance, will provide a more comprehensive view of software health and longevity. This research paves the way for significant advancements in software health management.